\documentclass[preprint,12pt,3p]{elsarticle}
\usepackage{graphicx}
\usepackage{geometry}
\usepackage[english]{babel}
\usepackage[T1]{fontenc}
\usepackage[utf8]{inputenc}
\usepackage{hyphenat}
\usepackage{booktabs}
\usepackage{gensymb}
\usepackage{siunitx}
\usepackage{color,soul}
\usepackage[dvipsnames]{xcolor}
\usepackage[version=4]{mhchem}
\usepackage{subfig}
\usepackage[figuresright]{rotating}
\usepackage{multirow}
\usepackage{setspace}
\usepackage{adjustbox}
\usepackage[sfdefault,condensed]{roboto} 
\usepackage[italic]{mathastext}          
\usepackage{enumitem}
\usepackage[pdftex]{hyperref}
\hypersetup{
  colorlinks   = true,
  urlcolor     = Orange,
  linkcolor    = RoyalBlue,
  citecolor    = RoyalBlue
}
\usepackage{cleveref}
\biboptions{sort&compress}

\begin{document}
    \begin{frontmatter}
        \journal{Acta Materialia}
        
        \title{\raggedright\LARGE\textbf{Recrystallisation phenomena in an ultrafine-grained Al--Mg--Si alloy revealed by correlative \textit{in situ} EBSD and TEM heating}}
        
        \author[GRAZA,GRAZB]{M. Theissing\corref{cor}}
        \ead{moritz.theissing@felmi-zfe.at}     
        \author[MUL]{S. Gonzaga}
        \author[MUL]{P. Willenshofer}
        \author[MUL]{T.M. Kremmer}
        \author[MUL]{S. Pogatscher}
        \author[GRAZA,GRAZB]{S. Mitsche}
        \author[BRA]{F.F. Sene}
        \author[MUL]{M.A. Tunes\corref{cor}}
        \ead{matheus.tunes@unileoben.ac.at}
        
        \cortext[cor]{Corresponding authors:}
        

        \address[GRAZA]{Institute of Electron Microscopy and Nanoanalysis, Graz University of Technology, Steyrergasse 17, 8010, Graz, Styria, Austria}
        
        \address[GRAZB]{Graz Centre for Electron Microscopy (FELMI-ZFE), Steyrergasse 17, 8010, Graz, Styria, Austria}
         
        \address[MUL]{Department of Metallurgy, Chair of Non-Ferrous Metallurgy, Montanuniversität Leoben, Franz-Josef-Strasse 18, 8700, Leoben, Styria, Austria}
              
        \address[BRA]{Institute of Aeronautics and Space, Brazilian Air Force, Praça Marechal Eduardo Gomes, 50, 12228-904, São José dos Campos, São Paulo, Brazil}
        
        \begin{abstract}
        \onehalfspacing
        \noindent Ultrafine-grained (UFG) aluminium alloys are promising lightweight structural materials for space applications, where a high grain-boundary density can act as sinks for irradiation-induced defects. Their deployment, however, is contingent on thermal stability: aluminium components in low-Earth orbit can reach $\sim$200~\textdegree C under solar irradiation, close to where severely deformed aluminium alloys recrystallise. Accurate, bulk-representative determination of recrystallisation onset is therefore essential, yet conventional \textit{in situ} transmission electron microscopy (TEM) heating is compromised by thin-film effects, ambiguous grain-boundary contrast, and small sampling volumes. Here, a UFG AA6061 (Al--Mg--Si) alloy produced by high-pressure torsion was studied by a first-of-its-kind comparison of \textit{in situ} TEM heating and \textit{in situ} electron backscatter diffraction (EBSD) heating, complemented by differential scanning calorimetry (DSC), analytical scanning transmission electron microscopy (STEM-EDX) and microhardness. \textit{In situ} EBSD sampled $\sim$$10^{3}$ grains from bulk material and resolved the microstructural evolution into sequential recovery, recrystallisation and grain-growth regimes, placing the onset of instability at $\sim$198~\textdegree C. Calorimetry, microhardness and nanoscale elemental mapping showed that grain refinement suppresses GP-zones formation and shifts precipitation to lower temperatures, with precipitation neither retarding recrystallisation nor restoring strength once the UFG structure is consumed. Revisiting the Brailsford--Bullough--Hayns sink-strength theory with a KAM-informed, temperature-dependent internal sink strength, we show that the grain-boundary sink strength collapses as recovery and recrystallisation proceed. We establish \textit{in situ} EBSD heating as a powerful \textit{in operando} methodology for the bulk-representative determination of microstructural instabilities in advanced metallic systems.
        \end{abstract}
        
    \end{frontmatter}

\cleardoublepage
{\tableofcontents}
\newpage

\onehalfspacing

\section{Introduction}
\label{sec:introduction}

\noindent Over the past three decades, the development of ultra-fine-grained (UFG) aluminium alloys via the methods of Severe Plastic Deformation (SPD) has become a central focus in metallurgy and materials science, stemming from the need to discover new functional materials \cite{prangnell2004ultra,raab2004continuous,valiev2006principles, valiev2006producing,langdon2007processing, liddicoat2010nanostructural,estrin2011ultrafine,langdon2013twenty, huang2013advances,xiang2019grain}. These alloys combine aluminium's inherent lightness with the benefits of both precipitation strengthening and exceptional hardening achieved through a combination of either UFG or nanostructured microstructures and appropriate heat-treatments. As a result, UFG aluminium alloys are promising candidates for advanced applications across diverse fields, including microelectromechanical systems (MEMS) and nanotechnology \cite{qiao2010fabrication}, cryogenic conditions \cite{harsha2021effect,chinh2021ultralow} as well as in high-temperature  \cite{he2020high}, corrosion environments \cite{esquivel2018excellent}, and inherent high mechanical loads \cite{ma2014aa7075}.

Space technology has recently been identified by our group as a promising field for the application of UFG aluminium alloys, owing to their potential mechanical properties and outstanding resistance to the extreme conditions of space \cite{tunes2020prototypic,willenshofer2022radiation,willenshofer2023precipitation,willenshofer2024comparative,tunes2025future}. In our solar system, radiation damage from solar energetic particles as well as thermal gradients present challenges for materials currently used or considered for satellites and spacecrafts \cite{baker2001satellite,holmes2002handbook,schwenn2006space,blasi2013origin,schwadron2018update,tunes2020prototypic,sznajder2023solar,ugwumadu2025effects,tunes2025future}. Coarse-grained (CG) commercial aluminium alloys -- such as the AA6061 within the Al--Mg--Si ternary system -- are known to exhibit poor radiation resistance to energetic particle bombardment, specially to highly energetic solar protons: MeV-range proton irradiations up to doses as low as 0.2 displacement-per-atoms (dpa) were reported to cause significant precipitate dissolution and mechanical properties' deterioration on the AA6061-T6 alloy \cite{lohmann1987microstructure}. Only few further studies have examined neutron irradiation effects on commercial aluminium alloys, mainly addressing AA2xxx, AA5xxx, AA6xxx, and AA7xxx series for low-temperature reactor applications \cite{vogl1965einfluss,katz1968precipitation,liu1972structural,farrell1981microstructure,piatti1984high,boning1987physics,ismail1990effect,ghauri2007effects,kolluri2016neutron}. Despite this limited research, it is evident that conventional aluminium alloys exhibit poor radiation resistance under extreme environments, particularly compared to conditions encountered in space dominated by solar particles and cosmic ray fluxes \cite{tunes2025future}. 

 Nanostructured and UFG metallic alloys -- those with grain sizes in the nanometre range -- have shown significant potential in mitigating the deleterious effects of radiation damage in extreme environments such as in space and thermonuclear fusion reactors. In these materials, a high density of grain boundaries promote actively sinks for radiation-induced point defects, thereby enhancing the alloys' resilience against the formation and evolution of inherent radiation-induced microstructural defects such as voids, dislocation loops, radiation-induced-segregation (RIS) and -precipitation (RIP) \cite{el2014ultrafine,el2014situ,el2017role,el2018nanohardness,zhang2018radiation,el2019outstanding,barr2019interplay,el2020revealing,el2020situ,el2020temperature,el2021helium,willenshofer2022radiation,el2023quinary}. Despite the high levels of radiation damage resistance that UFG and nanocrystalline materials can sustain, UFG aluminium alloys must demonstrate high thermal stability if they are candidate materials for application in space. Space components experience extreme temperature fluctuations, ranging from cryogenic levels in Earth's shadow to up to 150--180$^{\circ}$C in direct sunlight around Low-Earth Orbit (LEO) conditions \cite{Iyengar2022}. Similar metal-alloy plate temperature values are obtained from solar heat fluxes at 1 A.U. (Astronomical Unit) beyond the Van Allen radiation belts \cite{Juhasz2000}. Such thermal gradients require that UFG aluminium alloys exhibit thermal resilience and a stable microstructure, especially in preventing the phenomena of grain growth and recrystallisation.

Valiev and Langdon compiled findings from various studies employing \textit{in situ} heating with Transmission Electron Microscopy (TEM) to evaluate the thermal stability of new UFG aluminium alloys \cite{valiev2006principles}. These investigations revealed that most UFG aluminium alloys begin to recrystallise at temperatures as low as 200~\textdegree C \cite{valiev2006principles}, but due to known constraints of TEM instrumentation and the influence of thin film effects \cite{williams2009transmission}, the exact temperature at which microstructural instabilities begin in these alloys have not yet been established. Such early onset of recrystallisation suggests that the alloy may lose its initially engineered capacity to absorb radiation-induced defects at grain boundaries, jeopardising its intended application in environments where energetic particle irradiation is a major concern. In this context, the precise identification of the onset microstructural instabilities due to temperature is of paramount importance for the design of new UFG aluminium alloys intended for application in space.

In this study, a new UFG aluminium alloy was produced from a commercial CG AA6061 (Al--Mg--Si) base material using the SPD technique of High-Pressure Torsion (HPT). With the primarily major objective of accurately determining the onset of recrystallisation -- characterised by grain coarsening, coalescence, and recovery -- and to monitor other possible thermally induced microstructural changes, a first-time comparative investigation was conducted in this UFG AA6061 alloy using both conventional \textit{in situ} heating with TEM and the emerging technique of \textit{in situ} heating with Electron Backscatter Diffraction (EBSD) \cite{humphreys_insitu_1996,helbert_insitu_2012,adam_insitu_2017,chakkedath2018situ,he_insitu_2018,takajo_insitu_2019,adam_insitu_2021}. Complimentary Differential Scanning Calorimetry (DSC) and microhardness measurements were additionally employed to determine the role of precipitation in both the mechanical and thermal responses. Our experimental results show that \textit{in situ} heating with EBSD is as new powerful \textit{in operando} methodology to monitor real-time microstructural modifications in advanced UFG metallic alloy systems, allowing the precise estimation of the onset of recrystallisation in complex microstructures. By using these methods, we also demonstrate that for the UFG aluminium alloy 6061, precipitation is not able to retard or delay recrystallisation.

\section{Materials and methods}
\label{sec:matmet}

\subsection{Alloy synthesis with High-Pressure Torsion}
\label{sec:matmet:synthesishpt}
\noindent A commercially sourced AA6061 alloy was first remelted and cast into approximately 80 g laboratory-scale slabs using a vacuum induction furnace with a copper mould. The chemical composition of the cast material was determined via optical emission spectroscopy (OES) on a polished surface and this result is summarised in Table \ref{tab:chemical-composition}. Solution heat-treatment was performed in the remelted AA6061 alloy resulting in a microstructure that approximates of a supersaturated solid solution following procedures previously developed by Willenshofer \textit{et al.} \cite{willenshofer2022radiation,willenshofer2023precipitation,willenshofer4789331comparative}. To obtain a UFG AA6061 alloy, a cylindrical sample measuring 30 mm in diameter and 12 mm in height was prepared for SPD using High-Pressure Torsion (HPT). The process was performed at room temperature under a nominal pressure of 4 GPa for 10 revolutions to achieve a final UFG microstructure. As the radial position within the HPT disk determines the strain the material was exposed to, we ensured that samples for all microstructural analyses were taken from the same distance from the centre. These samples were taken in the disk plane and in the plane spanned between the rotation axis and radial direction. 

\begin{table}[hb!]
\centering
\caption{Elemental composition of the remelted AA6061 base alloy used in this work (wt.\%).}
\label{tab:chemical-composition}
\begin{tabular}{@{}llllllllll@{}}
\toprule
 & Si & Mg & Cu & Mn & Fe & Cr & Zn & Ti & Al \\ 
\midrule
AA6061 as-received & 0.77 & 0.86 & 0.23 & 0.13 & 0.43 & 0.21 & 0.07 & 0.04 & Balance \\ 
\bottomrule
\end{tabular}
\end{table}

\subsection{Scanning/Transmission Electron Microscopy}
\label{sec:matmet:STEM}
\noindent For Scanning/Transmission Electron Microscopy (S/TEM) sample preparation, alloy foils after HPT were ground and polished to a thickness of approximately 100 $\mu$m, and then 3 mm discs were subsequently punched. Electron-transparency of these discs was achieved by twin-jet electro-polishing in a nitric acid–methanol solution (volume ratio 1:3) at -30~\textdegree C and 13 V. S/TEM characterisation was performed using a Thermo Fisher Talos F200X electron microscope operating a Schottky-type field emission gun at 200 kV.

\subsection{Electron Backscatter Diffraction}
\label{sec:matmet:ESBD}
\noindent Slices with 1 mm thickness of the UFG AA6061 alloy were cut perpendicular from the HPT direction (same as for the TEM samples), ground, and polished to a grid size of 0.25 $\mu$m followed by electrochemical polishing with EP2 standard electrolyte (consisting of 140 ml distilled water, 800 ml ethanol, and 60 ml perchloric acid $60\%$) for 4 s at 5~\textdegree C and 26 V. Finally, the sample was mechanically polished in colloidal silica (40 nm particle size) for 3 h. EBSD was performed on a scanning electron microscope Zeiss Ultra55 equipped with an Thorlabs Fast Frame Rate Scientific camera and the EDAX OIM DC V7.3.1 data collection software for the EBSD measurements. For all EBSD measurements, a voltage of 20 kV and beam current of 40 nA was used. The stepsize for the overview maps of the material was set to 0.05 $\mathrm{\mu}$m. Spherical indexing \cite{lenthe2019spherical} was applied with the EMSphInx 0.2 software to the collected (\textit{in situ}) EBSD data in order to to bring the indexing success rate to a useable level. After re-indexing, the EBSD data was cleaned with grain dilation in the EDAX OIM Analysis V9.1 software, where around $2\%$ of the points were changed. Following processing of the EBSD data was performed with the EDAX OIM Analysis V9.1 and the MTEX toolbox \cite{bachmann2010texture} in MATLAB.

\subsection{\textit{In situ} TEM and \textit{in situ} EBSD heating experiments}
\label{sec:matmet:insitu}
\noindent To track the microstructural stability and onset recrystallisation of the UFG AA6061 alloy, heating experiments involving both \textit{in situ} TEM and \textit{in situ} EBSD were performed. \textit{In situ} TEM heating was carried out with a Protochips Fusion Select \textit{in situ} heating/cooling holder with an uncoated MEMS e-chip. Here, the samples were subjected to a linear heating rate of 10~\textdegree C$\cdot \mathrm{min}^{-1}$. \textit{In situ} EBSD was performed in a PID controlled CH0-4 Kammrath and Weiss heating stage. Here, the samples were subjected in one experiment to two linear heating rates: 1~\textdegree C$\cdot\mathrm{min}^{-1}$ and 10~\textdegree C$\cdot \mathrm{min}^{-1}$. In the latter experiment, heating was followed by holding at the starting temperature of recrystallisation for several hours. For the \textit{in situ} EBSD a scan size of $15x15\mu \mathrm{m}^2$ and a step-size of $0.1\mu\mathrm{m}$ was used.

\subsection{Differential Scanning Calorimetry}
\label{sec:matmet:DSC}
\noindent In addition to \textit{in situ} TEM and \textit{in situ} EBSD heating experiments, this study also employed DSC to investigate the thermal response of the UFG AA6061. The technique involves monitoring the heat flux between a sample and a reference material (pure aluminium) as they are heated at a constant rate in a controlled environment. As phase transitions occur (such as melting or precipitation), the sample absorbs or releases energy in the form of heat, which is compared directly to known enthalpy changes determined from previous experiments and literature. In this work, following a protocol established in a previous work \cite{willenshofer2023precipitation}, the Netzsch 204DSC F1 Phönix device was employed for the DSC measurements with molecular nitrogen serving as a purge and protective gas at a flow rate of 20 ml$\cdot$min$^{-1}$. We also performed DSC measurements in the CG version of the AA6061 alloy (\textit{i.e.} its commercial variant). We report herein results obtained with DSC in both CG and UFG AA6061 alloy conditions. The heating rate for all thermal analysis experiments were performed at the same heating rate as for the \textit{in situ} TEM and EBSD heating experiments: 10~\textdegree C$\cdot\mathrm{min}^{-1}$. The DSC measurements were carried out three times with both the CG and UFG AA6061 alloys for reproducibility and statistics.

\subsection{Microhardness}
\label{sec:matmet:microhardness}
\noindent Before the microindentation and microhardness measurements, the UFG and CG AA6061 groups of the aluminium alloy samples were subjected to the thermal treatment during 60 min at temperatures defined by the experimental peaks detected by the DSC analysis. These heat-treatments were performed by using an electrical box furnace in air atmosphere. The UFG AA6061 samples were heat-treated at 225 and 316~\textdegree C whilst the CG AA6061 samples were heat-treated at 88, 242, 298, and 477~\textdegree C.

The microindentation was performed by using by a standard hardness test device with a typical indentation depth about 20 to 50 $\mu$m. The diagonal values were measured in at least 10 different regions on both the transverse and the longitudinal cross-sections.  The Vickers microhardness was determining for the samples after the thermal treatment, by using the primary equation:
\begin{equation}
\label{eq:1}
	H_v = 1854.4 \frac{P}{d^2}
\end{equation}

In the equation \ref{eq:1}, $P$ is the applied load in kilogram-force [kgf] and $d$ is the length of the indentation diagonal in millimetre [mm]. The dwell time was set to 12 s and the applied load mass used was 50 g for all investigated samples.

\section{Results and discussion}
\label{sec:res}

\subsection{Characterisation of the UFG AA6061 after HPT}
\label{sec:resdis:pristine}
\begin{figure}[ht!]
	\centering
	\includegraphics[width=0.9\textwidth]{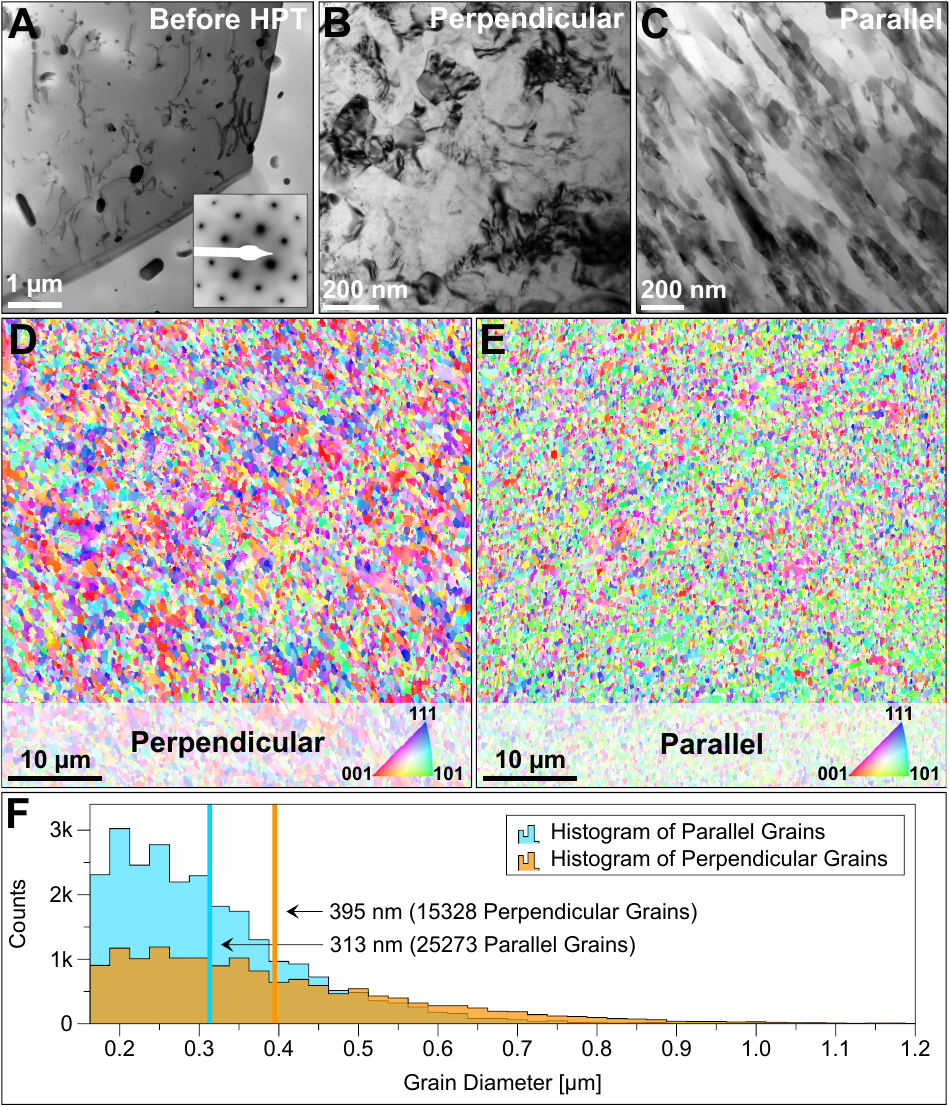}
	\caption{\textbf{Characterisation of the UFG AA6061 alloy before and after HPT} | \textbf{A} BFTEM micrograph of the CG commercial-grade AA6061 alloy as-received. \textbf{B} BFTEM micrograph of the UFG AA6061 alloy after deformation with grains perpendicular to HPT direction.  \textbf{C} BFTEM micrograph of the UFG AA6061 alloy after deformation with grains parallel to HPT direction. \textbf{D} EBSD IPF map of the UFG AA6061 alloy after deformation perpendicular to HPT direction. \textbf{E} EBSD IPF map of the UFG AA6061 alloy after deformation parallel to HPT direction. \textbf{F} Grain diameter histograms from data in \textbf{D} and \textbf{E}. Note: Error bars (representing the standard error of the mean) in the average grain sizes in \textbf{F} were suppressed due to the high number of counted grains.}
	\label{fig:01}
\end{figure}

\noindent Before performing the \textit{in situ} TEM and EBSD experiments, a complete characterisation of the UFG AA6061 alloy is necessary. The microstructure of the AA6061 alloy before and after HPT is shown in the Bright-Field TEM (BFTEM) micrographs in Figs. \ref{fig:01}A-C. Fig. \ref{fig:01}A features a Selected-Area Diffraction (SAED) as inset that was taken along the [001]$_{Al(FCC)}$ zone axis to show that the alloy precursor to HPT was in the solution heat-treated state. Here, no super-lattice and/or satellite spots resembling hardening phases can be noted, as expected for this temper. Although micrometre-sized dispersoids and/or secondary phases are noted within the microstructure of the AA6061 alloy before HPT, these phases do not preclude SPD as they pose neither notable nor significant hardening to the alloy. Figs. \ref{fig:01}B-C indicate that the grain morphology of the UFG AA6061 alloy was discovered to be dependent on the specimens' cutting direction, \textit{i.e.}, along the perpendicular direction, the grains were observed to be more equiaxed whereas along the parallel direction, a more elongated shape was noticeable. Complimentary EBSD assessment of the UFG AA6061 alloy is shown in the Inverse Pole Figures (IPF) on Figs. \ref{fig:01}D and \ref{fig:01}E for the perpendicular and parallel directions, respectively. The IPFs show preferred orientations for the two directions: in the perpendicular sample the [001] and [111] orientations are dominant whereas in the parallel sample the [101] orientation is dominant. Average grain size (equivalent circular diameter) as a function of the cutting direction is shown in the histograms in Fig. \ref{fig:01}F obtained from the orientation data with a grain boundary angle set to $5\,^{\circ}$. Following the latter procedure, along the perpendicular direction the average grain size and standard deviation were estimated to be 395$\pm$19 nm whereas on the parallel direction these values were 313$\pm$12 nm. 

A combined TEM and EBSD analysis of AA6061 after HPT is essential for understanding the evolution of this microstructure during heating. It is worth noting that modern EBSD techniques with high spatial resolution -- and the recently introduced spherical indexing \cite{lenthe2019spherical} -- enable a combination of fast data acquisition and high resolution. This allows for a grain size evaluation for such ultrafine-grained microstructures and, therefore, cover of 15000-25000 individual grains, thus generating statistically validated results. This is an exceptional capability for tracking microstructural evolution of new alloys, especially to monitor modifications in the alloy's microstructure posed by heating.

\subsection{\textit{In situ} microstructural response to heating via TEM}
\label{sec:res:insituTEM}
\begin{figure}[ht!]
	\centering
	\includegraphics[width=0.85\textwidth]{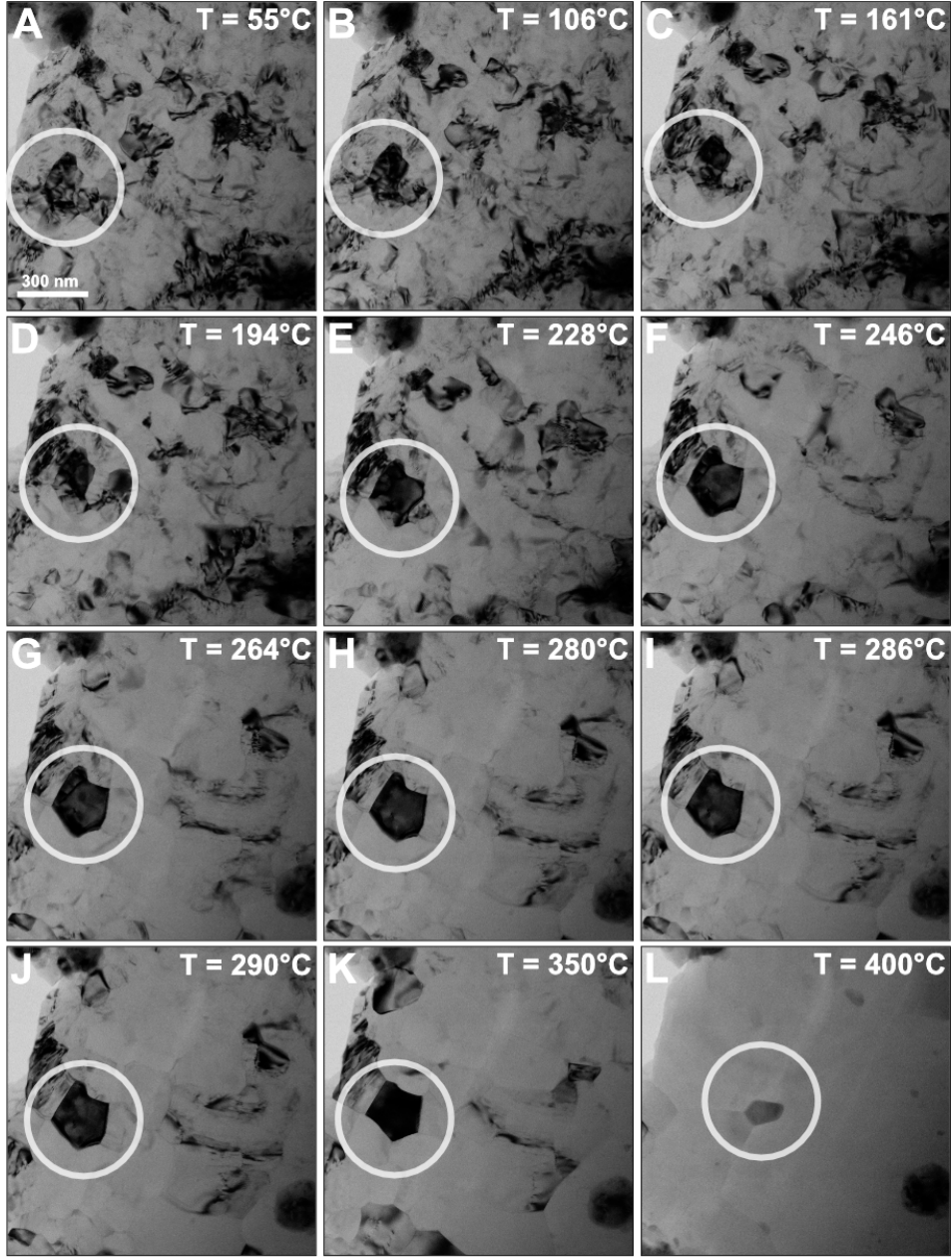}
	\caption{{{\textit{In situ} TEM with MEMS heating experiment | \textbf{A} to \textbf{L} show the evolution of grains during heating with 10~\textdegree C$\cdot \mathrm{min}^{-1}$ measured with BFTEM parallel to HPT direction. The white circles in \textbf{D},\textbf{E} and \textbf{F} indicate an area where recrystallisation starts. Note: the scale bar in \textbf{A} is applied to all micrographs in the figure; the micrographs exposure levels were increased by 25\% for better visualisation.}}}
	\label{fig:02}
\end{figure}

\noindent \textit{In situ} TEM with MEMS heating reveals how the UFG AA6061 alloy microstructure changes during heating with a rate of 10~\textdegree C$\cdot \mathrm{min}^{-1}$ as shown in Figs.\ref{fig:02}A-L. 

It is worth noting that this experiment was performed with the sample parallel to HPT direction. All of these micrographs were captured with BFTEM to allow the visualisation of the grains and their boundaries. Fig.\ref{fig:02}A depicts the initial deformed microstructure recorded at 56~\textdegree C. The higher temperature starting point was chosen to point out that the \textit{in situ} TEM heating holder with MEMS is more stable as no change due to recovery and/or recrystallisation is expected at this temperature when compared to room temperature. Up to a temperature of 194~\textdegree C (Fig.\ref{fig:02}D) no major change in the microstructure is noticeable. Above 194~\textdegree C, as marked with white circles in Figs.\ref{fig:02}E-F, one can observe the very first noticiable change in the morphology of the grains. This microstructural tracking under controlled heating allows for the estimation of the recrystallisation temperature between 194~\textdegree C and 246~\textdegree C, but given the fast dynamics of \textit{in situ} TEM heating, a more precise temperature at the onset of recrystallisation was not possible to obtain. After 246~\textdegree C, one can note new grains forming and/or pre-existing grain growing via coalescence during the heating ramp. This latter is particularly noticeable in Figs.\ref{fig:02}G-K. At the final temperature of 400~\textdegree C, recrystallisation is completed and a few larger grains are observed in the micrograph Figs.\ref{fig:02}L. 

While \textit{in situ} TEM heating presents a viable route for monitoring microstructural recrystallisation and grain growth, some challenges should be noted. The most significant limitations include (i) the constrained number of grains that can be monitored and analysed (even considering lower magnifications), (ii) the thin film effect that has been reported in acceleration of thermal phenomena in electron-transparent metals \cite{singh2017effects,van2020size,coradini2024unravelling}, and (iii) the absence of a reliable method to either detect or define the onset of microstructural alterations at determined/accurate temperatures. In the domain of nanometallurgy, the thin film effect is reported to cause a shift of the onset of recrystallisation to lower temperatures \cite{coradini2024unravelling}. Furthermore, the complex contrast-forming mechanisms resulting from electron beam interactions with electron-transparent samples can lead to ambiguous grain boundary differentiation, adding an extra layer of complexity to the experiments and their analysis. The contrast-forming mechanisms within TEMs do not allow a fast temporal differentiation of recovery and recrystallisation at the start of the transformation process. All these challenges are in-line with recent literature on the limitations of thermal stability and heating experiments within \textit{in situ} TEM \cite{el2021limitations}.

\subsection{\textit{In situ} microstructural response to heating via EBSD}
\label{sec:res:insituEBSD}
\noindent In parallel to the \textit{in situ} TEM heating with MEMS, complimentary experiments using the emerging technique of \textit{in situ} EBSD with heating were carried out on pristine UFG AA6061 specimens (perpendicular to HPT direction). For these experiments \textit{in situ} EBSD heating experiments, the same heating rate of 10~\textdegree C$\cdot\mathrm{min}^{-1}$ was used, but after the onset of recrystallisation, the active heating ramp was stopped and the temperature was held isothermally. 

\begin{figure}
	\centering
	\includegraphics[width=\textwidth]{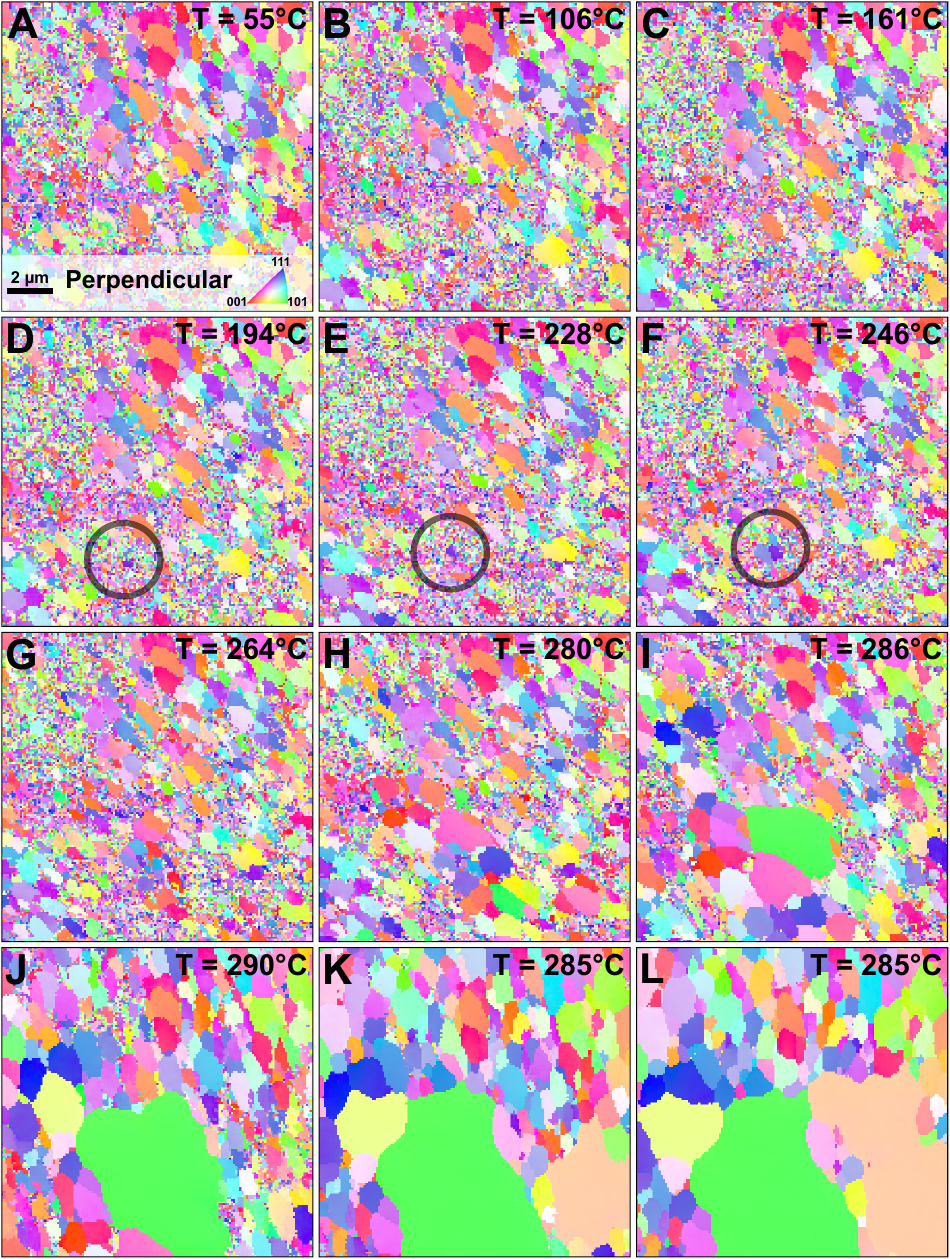}
	\caption{\textbf{\textit{In situ} EBSD with heating experiment} | \textbf{A} to \textbf{L} show the evolution of grains during heating with 10~\textdegree C$\cdot \mathrm{min}^{-1}$ (\textbf{A} to \textbf{H}) and holding at 285~\textdegree C (\textbf{I} to \textbf{L}) after recrystallisation starts measured with EBSD {perpendicular to HPT direction}. The black circles in \textbf{D},\textbf{E} and \textbf{F} indicate an area where recrystallisation starts. The isothermal holding time were 0 min for \textbf{I}, 2 min for \textbf{J}, 2 h for \textbf{K} and 19 h for \textbf{L}. Note: the scale bar and IPF legend in \textbf{A} is applied to all micrographs in the figure.}
	\label{fig:03}
\end{figure}

Figs.\ref{fig:03}A-L show the result of \textit{in situ} EBSD with heating in terms of IPF maps. In the first 5 maps (Figs.\ref{fig:03}A-E) and up to a temperature of 228~\textdegree C, the microstructure of the UFG AA6061 alloy remains resilient to temperature-induced changes in recovery. The noise in the left part of the EBSD maps suggests the presence of smaller grains with greater deformation.  However, the higher acquisition speed used during the \textit{in situ} experiment impairs their proper visualisation. After reaching a temperature of 246~\textdegree C, a very first recrystallised grain is denoted in the black circle in Figs.\ref{fig:03}D-F. This result can be interpreted that the onset recrystallisation lies between 228~\textdegree C and 246~\textdegree C. After the scan in Fig.\ref{fig:03}I at 286~\textdegree C, recrystallisation becomes significant and affects the whole alloy's microstructure. At this point, the active heating rate was shut off and the temperature was held. In the isothermal part of the experiment (Figs.\ref{fig:03}I-L) we see significant growth to large grains via coalescence. Comparison of grain shapes before and after recrystallisation reveals a significant change in the microstructure of the original alloy after HPT. Prior stopping the heating ramp isothermally, the grains were more distorted, likely due to some thermal drift caused by the high heating rate and small scan window (15x15 $\mu\mathrm{m}^2$) over a one-minute period. By switching to an isothermal state, the thermal drift markedly reduced, resulting in less grain distortion. It is worth noting that this slight morphological distortion likely caused by thermal drifting does not impact the results of our analysis. A one-minute scan spans a temperature interval of 10~\textdegree C; the midpoint of this interval was therefore taken as the scan temperature. Although the transformation proceeds continuously during each scan, this does not affect the analysis, as any changes in the microstructure are evaluated between successive scans rather than within a single scan. This, however, may slightly reduce the temperature precision. In contrast to thin-film effects observed in TEM, surface effects in EBSD -- such as altered kinetics at the specimen surface -- are unlikely. Nucleation was observed at the surface as well as grain growth from the bulk towards the surface. Furthermore, post-experiment inspection revealed no alteration of the surface quality compared to the pre-experiment state, indicating the absence of significant surface effects. 

\subsection{Impact of heat-treatment on the microstructure and mechanical properties}
\label{sec:res:microhardness_and_DSC}
\noindent The AA6xxx series alloys are considered age-hardenable \cite{pogatscher2011mechanisms}, thus precipitation is expected to occur in both CG and UFG AA6061 alloy conditions as a function of temper. However, the \textit{in situ} TEM and EBSD heating experiments on the UFG AA6061 after HPT presented in both Figs. \ref{fig:02} and \ref{fig:03} indicate that neither nanoscale nor microscale precipitation was noticeable within its microstructure solely due to temperature ramping at the chosen conditions. In order to investigate if precipitation phenomena somehow affects the microstructure, and consequently the recrystallisation behaviour of the newly synthesised UFG AA6061 alloy, an additional study with DSC, microhardness and STEM-EDX characterisation was performed after heat-treatment at different temperatures. It is worth emphasising that, as stated in the materials and methods subsection \ref{sec:matmet:DSC}, both the \textit{in situ} TEM/EBSD heating and the DSC experiments were performed with the same heating rate of 10~\textdegree C$\cdot\mathrm{min}^{-1}$.

\begin{figure}[ht!]
\centering
\includegraphics[width=\textwidth]{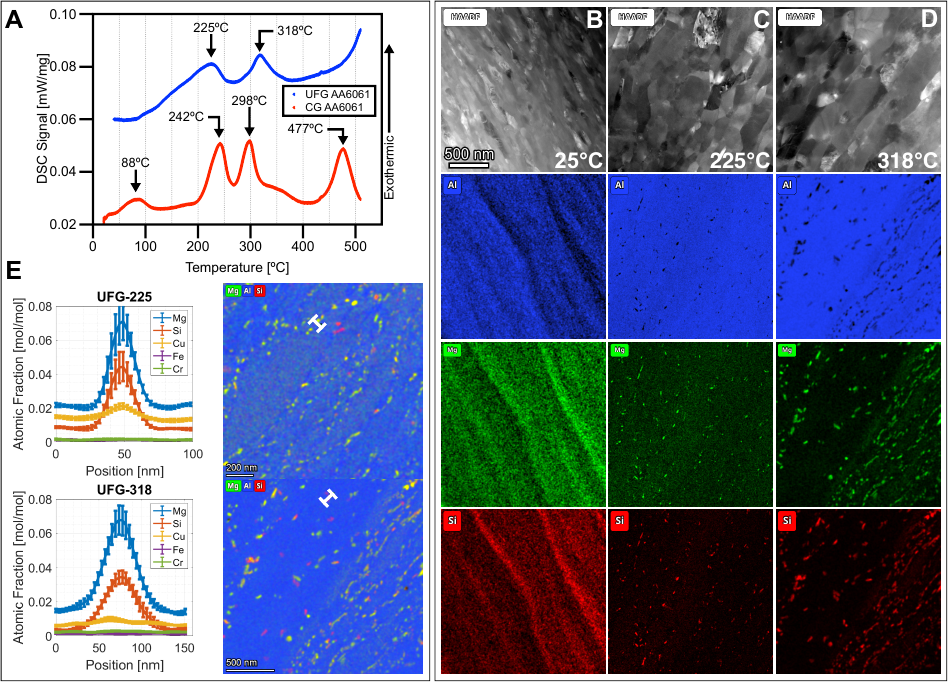}
\caption{\textbf{Effects of heat-treatment in the microstructure of the studied alloys} | In \textbf{A} the DSC curves for both the CG (red) and UFG (blue) AA6061 alloys are presented. The HAADF and STEM-EDX maps for the UFG AA6061 alloy microstructures after heat-treatments at \textbf{B} 25~\textdegree C, \textbf{C} 225~\textdegree C, and \textbf{D} 318~\textdegree C are shown. In \textbf{E}, line-profile chemical analysis of precipitates identified in the UFG AA6061 alloy at both 225~\textdegree C and 318~\textdegree C are shown with the quantification and Mg/Si ratio presented in Table \ref{tab:quantificationUFG}. Note: In the quantification presented in \textbf{E} the elements Cu, Fe, and Cr are traces and within the limit of detection.}
\label{fig:04}
\end{figure}

The precipitation behaviour of CG AA6061 alloy as analysed by the DSC is shown in Fig. \ref{fig:04}A. The heat-flow trace reveals four distinct exothermic reactions occurring at approximately 88~\textdegree C, 242~\textdegree C, 298~\textdegree C, and 477 ~\textdegree C. The observed precipitation sequence is in agreement with trends reported in earlier calorimetric investigations of AA6061 alloys \cite{dutta1991calorimetric, edwards1998precipitation, yassar2005transmission}. The low-temperature exothermic event at 88~\textdegree C is associated with the formation of solute clusters, including Si-rich clusters, Mg-rich clusters, mixed Mg–Si co-clusters, as well as GP-I zones \cite{dumitraschkewitz2018clustering}. The reaction centred at 242~\textdegree C corresponds to the precipitation of the metastable $\beta^{\prime\prime}$ phase, whereas the peak at 298~\textdegree C is attributed to the subsequent formation of $\beta^{\prime}$-phase. The high-temperature peak at 477~\textdegree C reflects the precipitation of the equilibrium $\beta$-phase (Mg$_2$Si) \cite{dutta1991calorimetric, edwards1998precipitation, yassar2005transmission}. In contrast to the analyses reported by Dutta \textit{et al.} \cite{dutta1991calorimetric} and Edwards \textit{et al.} \cite{edwards1998precipitation}, no peak deconvolution was applied to the second exothermic signal in the present study. Specifically, the DSC curve in Fig. \ref{fig:04}A does not exhibit indications of overlapping reactions associated with GP-I, GP-II, and $\beta^{\prime\prime}$-phase precipitation, as previously reported by Dutta \textit{et al.}\cite{dutta1991calorimetric}. Consequently, a single, well-defined peak description was considered sufficient for the herein investigated CG AA6061 alloy.

Similar to the CG AA6061 commercial alloy, the precipitation behaviour of the UFG AA6061 alloy was also examined DSC with the resulting heat-flow responses shown in Fig. \ref{fig:04}B. In contrast to the CG AA6061 alloy, which exhibited four distinct exothermic reactions, the UFG alloy is characterised by presence of only two exothermic peaks, located at approximately 225~\textdegree C and 318~\textdegree C, thus requiring further electron microscopy characterisation amid the microstructural identification of precipitates.

STEM–EDX mapping was performed for the elemental identification and quantitative analysis of the UFG AA6061 alloy microstructure at temperatures corresponding to the DSC exothermic events. In addition, a reference specimen in the as-received condition was examined. Elemental distribution maps of the principal alloying elements (Al, Mg, and Si) are shown in Figs. \ref{fig:04}B–D. While no precipitate formation is detected at 25~\textdegree C (Fig. \ref{fig:04}B), distinct Mg-Si-rich precipitates are clearly observed after heating to 225~\textdegree C and 318~\textdegree C, respectively shown in Fig. \ref{fig:04}C and \ref{fig:04}D. At 225~\textdegree C, precipitation occurs predominantly within the grain interiors (transgranular) and within the grain boundaries (intragranular). Upon further heating to 318~\textdegree C, these precipitates increase in size and number and remain largely at intragranular positions. Based on the STEM–EDX elemental maps from Figs. \ref{fig:04}B–D, the Mg/Si atomic percentage ratio and the characteristic size of the precipitates could be assessed for the thermal conditions at which precipitation was detected, namely at 225~\textdegree C and 318~\textdegree C. The corresponding quantitative results are summarised in Fig. \ref{fig:04}E and Table \ref{tab:quantificationUFG}. Precipitate dimensions were evaluated under the assumption of a spherical morphology. Consistent with the observations discussed above, the specimen analysed in the as-received condition did not exhibit any hardening precipitates, nor evidence of solute clusters or GP zones.

\begin{table}[!th]
\caption{Mg/Si ratio and average size measurements of the precipitates identified in the UFG AA6061 alloy.}
\centering
\begin{tabular}{|c|c|c|c|c|c|c|c|c|}
\hline
\textbf{Temperature Peak} & \textbf{225~\textdegree C} & \textbf{318~\textdegree C} \\
\hline
Mg/Si ratio (Radius) & 1.60 (43.3nm) & 1.96 (73.5nm) \\ 
\hline
\end{tabular}
\\ \tiny{Note: An average of 20 precipitates have been quantified for each case presented.}
\label{tab:quantificationUFG}
\end{table}

The combined DSC and STEM-EDX results (Figs. \ref{fig:04}A and \ref{fig:04}B-E) describe the temperature-dependent microstructural evolution of the UFG AA6061 alloy. No GP-zone formation was detected in the UFG condition, as evidenced by the absence of DSC peaks below 225~\textdegree C and the lack of corresponding features in the STEM-EDX maps. As precipitation is diffusion-controlled and strongly influenced by grain size, the modified precipitation sequence relative to the CG alloy is attributed to accelerated diffusion and enhanced precipitation kinetics in the UFG microstructure, which is in-line with the recent literature of UFG aluminium alloys \cite{sha2014strength,willenshofer4789331comparative}. The high grain-boundary density provides fast diffusion paths, which may suppress GP-zone stability and promote an early transformation from $\beta^{\prime\prime}$ to $\beta^{\prime}$ or $\beta$ phases \cite{sha2014strength}.

In comparison to the CG alloy, precipitation-related DSC peaks are shifted to lower temperatures in the UFG state, from 242~\textdegree C to 225~\textdegree C and from 477~\textdegree C to 318~\textdegree C. Consequently, precipitate size and Mg/Si ratio were analysed at these two temperatures. At 225~\textdegree C, the estimated Mg/Si ratio of 1.6 is indicative of the $\beta^{\prime}$ phase \cite{andersen1998crystal}. At 318~\textdegree C, the precipitates were larger and more rounded, with a transition towards partially transgranular precipitation. The corresponding Mg/Si ratio of 1.96 suggests the formation of the equilibrium $\beta$ (Mg$_2$Si) phase \cite{andersen1998crystal}.

\begin{figure}[hb!]
\centering
\includegraphics[width=0.55\textwidth]{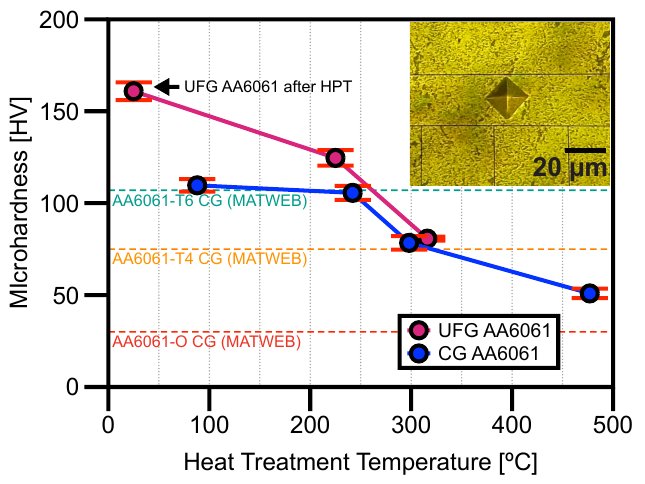}
\caption{\textbf{Hardness response versus heat-treatment} | Microhardness versus the heat-treatment temperatures for both CG and UFG AA6061-T6 alloys, where the vertical dashed lines show standard Vickers hardness values for the commercial CG AA6061 alloy at solution annealed (O), T4 and T6 (peak-hardness) tempers \cite{MatWeb_6061T6}.}
\label{fig:05}
\end{figure}

Both CG and UFG AA6061 alloys were heat treated at the characteristic DSC peak temperatures, followed by microindentation testing; the results are shown in Fig. \ref{fig:05}. In both cases, hardness decreases with increasing heat-treatment temperature. However, the UFG alloy exhibits a markedly steeper reduction in microhardness between the as-HPT condition and the first DSC peak at 242~\textdegree C. As discussed in subsection \ref{sec:res:insituEBSD}, the onset of recrystallisation in the UFG AA6061 alloy occurs between 228~\textdegree C and 246~\textdegree C. This indicates that the initial strength imparted by HPT (\textit{i.e.} an ultrafine-grained microstructre) is rapidly lost due to recrystallisation, with precipitation neither inhibiting recrystallisation nor providing a significant hardening contribution after recrystallisation: the latter further confirmed with the microindentation testing after heat-treatment at 318~\textdegree C.

\section{Overall discussion}
\label{sec:dis}

\subsection{Accurate determination of the recrystallisation temperature}
\label{sec:dis:determination}
\noindent \textit{In situ} EBSD with heating offers distinct advantages compared to \textit{in situ} TEM with MEMS heating, including the ability to monitor larger number of grains, the absence of the pronounced thin film effect as experiments are performed on bulk material, and enhanced visualisation of grain boundaries through inherent EBSD contrast. In this work, we have shown that \textit{in situ} EBSD enables the determination of the recrystallisation temperature with even increased precision when compared with \textit{in situ} TEM. Due to the relatively high heating rate of 10~\textdegree C/min presented in the previous section, the temporal resolution for determining the onset of recrystallization using \textit{in situ} EBSD was limited. Therefore, a second experiment was conducted at a lower heating rate of 1~\textdegree C/min. The reduced heating rate allowed a substantially larger number of \textit{in situ} scans to be acquired, resulting in improved temporal resolution.

\begin{figure}[hb!]
\centering
\includegraphics[width=\textwidth]{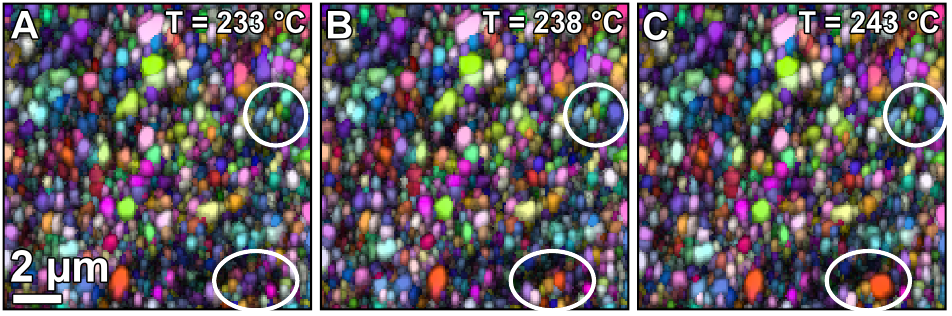}
\caption{\textbf{Estimation of the onset of recrystallization} | \textbf{A} to \textbf{C} show the first nucleation of recrystallized grains collected during an \textit{in situ} EBSD experiment with a heating rate of 1~\textdegree C/min. The white ellipse and circle highlights regions where new grains emerge, or existing (sub-) grains become nuclei and start to grow. Note: the data representation is IPFs overlayed on the confidence index after spherical indexing; the scale bar in \textbf{A} applies to \textbf{B} and \textbf{C}.}
\label{fig:Moritz_1}
\end{figure}

Figure \ref{fig:Moritz_1} presents three representative scans from the 1~\textdegree C/min heating experiment, capturing the initial stages of recrystallization or, generally, the earliest microstructural changes experience by the UFG AA6061 alloy. The IPF maps are overlaid with the confidence index (CI), which facilitates the identification of (any) microstructural changes. After spherical indexing, the CI behaves similarly to image quality (IQ) or band contrast (bc) maps: brighter regions correspond to areas where indexing quality is higher. In general, regions with a high degree of deformation or elevated defect density tend to exhibit lower indexing quality. As the UFG AA6061 ally herein investigated is severely deformed and, therefore, characterized by an overall high defect density, any local increase in CI are most likely associated with recovery and/or recrystallization processes \cite{wright2006ebsd}. Comparing Figure \ref{fig:Moritz_1}A and Figure \ref{fig:Moritz_1}B, a new red grain becomes visible within the white ellipse. This grain subsequently grows, as shown in Figure \ref{fig:Moritz_1}C, providing clear evidence of recrystallization. In addition, the grain highlighted by the white circle becomes noticeably brighter as the temperature increases from 233~\textdegree C to 238~\textdegree C. This increase in CI suggests a reduction in defect density within the grain, indicating that recovery and/or recrystallization has occurred.

The complete recrystallization process observed during the second \textit{in situ} EBSD heating experiment is shown in Figure S1 of the supplemental information. The final microstructures obtained after both heating experiments are visually comparable, consisting of a mixture of several large grains and a high number of small grains (Figure S2, supplemental information). However, the grain size distributions reveal some differences: the fraction of small grains (grain diameter $<$ 1 $\mu$m) is higher after heating at 1~\textdegree C/min, whereas heating at 10~\textdegree C/min shows more intermediate-sized grains with diameters up to approximately 2.5 $\mu$m. The fraction of larger grains is similar in both conditions. The mean grain size after the in situ experiments is 1.00 $\mu$m for heating at 10~\textdegree C/min and 0.92 $\mu$m for heating at 1~\textdegree C/min.

Nevertheless, visual inspection of the \textit{in situ} EBSD heating data provides valuable insight into the microstructural evolution and the onset of recrystallization of UFG alloys. To quantify the transformation process, the evolution of the average grain size with time is plotted in Figure 2 for both experiments. The corresponding temperature profiles are included to facilitate correlation between microstructural changes and thermal conditions.

\begin{figure}[hb!]
\centering
\includegraphics[width=1\textwidth]{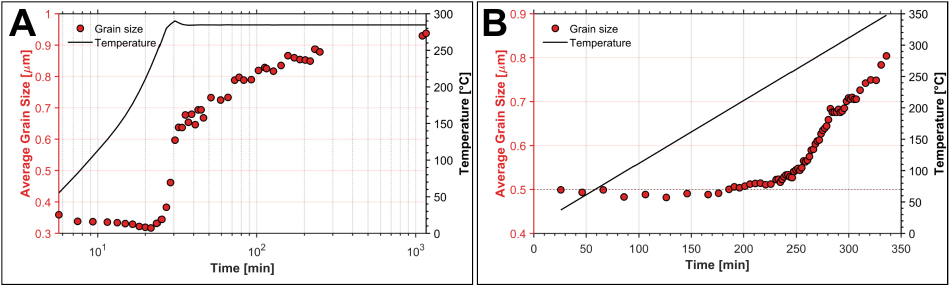}
\caption{\textbf{Tracking evolution of grain size during \textit{in situ} EBSD heating} | \textbf{A} shows the evolution average grain size during heating with 10~\textdegree C/min and holding with 286~\textdegree C over time. \textbf{B} shows the evolution average grain size during heating with 1~\textdegree C/min over time. Note: Due to the long measurement time the x-axis in \textbf{A} is in logarithmic scaling. In both graphs the temperature curve is included.}
\label{fig:Moritz_2}
\end{figure}

Figure \ref{fig:Moritz_2}A,B shows the grain size evolution during annealing at a heating rate of 10~\textdegree C/min up to 286~\textdegree C, followed by an isothermal hold of approximately 18 h. Initially, the average grain size decreases until a turning point is reached at around 228~\textdegree C. This apparent grain refinement is most likely a measurement artifact caused by the combination of the high heating rate and elevated temperatures, both of which reduce EBSD pattern quality and can affect grain reconstruction. Beyond this temperature, the average grain size increases rapidly over a short period, indicating the onset and progression of recrystallization following nucleation. After the transition to the isothermal holding stage, the slope of the grain size curve decreases noticeably. This behaviour can likely be attributed to two factors. First, the rate of energy input decreases once heating up ceases, reducing the driving force available for further recrystallization. Second, the transformation gradually shifts from recrystallization-dominated kinetics to grain growth, which generally proceeds at a slower rate.

The experiment conducted at 1~\textdegree C/min (Figure \ref{fig:Moritz_2}) reveals a more detailed picture of the transformation process due to its higher temporal resolution. Four distinct stages can be identified. In the first stage, up to approximately 201~\textdegree C, the average grain size remains essentially constant. In the second stage, between 201~\textdegree C and 247~\textdegree C, a small jump and then nearly constant increase in grain size is observed. This behaviour is consistent with recovery-related processes, including subgrain growth and the early stages of recrystallization nucleation. The third stage is characterized by an initially slow, followed by a rapid increase in average grain size, reflecting the progression of recrystallization and the growth of newly formed grains. Finally, above approximately 296~\textdegree C, the rate of grain growth increase decreases and becomes nearly constant. This fourth stage is interpreted as grain growth after recrystallization has largely been completed.

\subsection{Unravelling the physical mechanisms during annealing of UFG aluminium alloys}
\label{sec:dis:stress_annhilation}

\begin{figure}[hb!]
\centering
\includegraphics[width=1\textwidth]{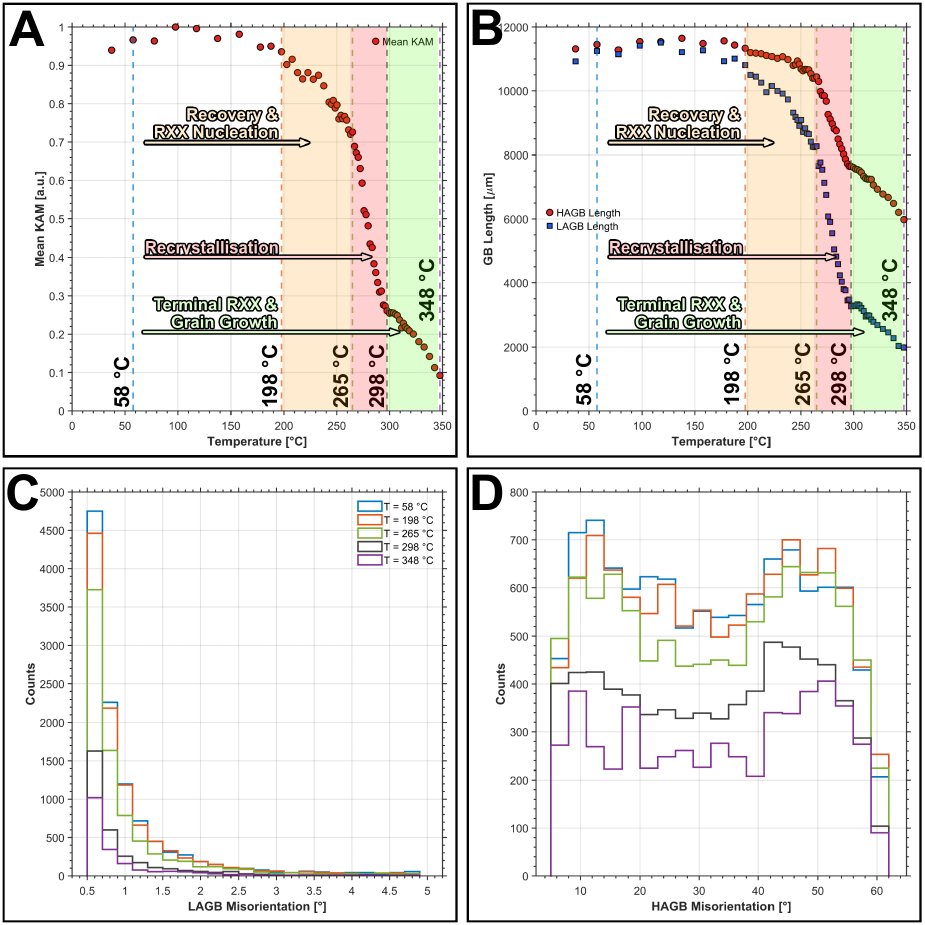}
\caption{\textbf{Evaluation of the annealing mechanism after heating with 1~\textdegree C/min} | \textbf{A} shows evolution of the normalized average KAM over temperature. \textbf{B} shows the grain boundary length of low angle grain boundaries (LAGB) and high angle grain boundaries (HAGB) over temperature. \textbf{C} shows histograms of the misorientation angle of LAGB at selected temperatures. \textbf{D} shows histograms of the misorientation angle of HAGB at selected temperatures. Note the coloured vertical dashed lines in \textbf{A} and \textbf{B} corresponds to the evaluated temperatures in \textbf{C} and \textbf{D}. The legend of \textbf{C} is also applied to \textbf{D}.}
\label{fig:Moritz_3}
\end{figure}

\noindent A more detailed analysis of the 1~\textdegree C/min \textit{in situ} EBSD heating experiment allows the different mechanisms active during annealing to be distinguished. The corresponding results are presented in Figure \ref{fig:Moritz_3}. Figure \ref{fig:Moritz_3}A shows the evolution of the normalized average kernel average misorientation (KAM) as a function of temperature with following equation:

\begin{equation}
\mathrm{KAM_{norm}}(T)=\frac{\mathrm{KAM}(T)-\min(\mathrm{KAM})}{\max(\mathrm{KAM})-\min(\mathrm{KAM})}
\label{eq:mean_KAM}
\end{equation}

The KAM is a parameter for monitoring microstructural changes as it correlates with the stored energy (\textit{e.g.}, in this case plastic deformation) in the material and, consequently, with the local defect density \cite{godfrey2015characterization}. A reduction in the normalized average KAM is interpreted as a decrease in stored energy and, hence, softening of the material. Up to approximately 198~\textdegree C, no significant change in the normalized average KAM is observed. The small variations present in this temperature range are attributed to experimental scatter resulting from the rapid EBSD acquisition required for the \textit{in situ} measurements. Between 198~\textdegree C and 265~\textdegree C, a moderate decrease in the normalized average KAM is evident. This behaviour can be attributed primarily to recovery processes and the nucleation stage of recrystallisation, both of which reduce the defect density and stored energy within the microstructure. A much steeper decline in normalized average KAM is observed between 265~\textdegree C and 298~\textdegree C. This temperature range corresponds to the stage in which recrystallization becomes the dominant softening mechanism, leading to a rapid reduction in stored energy as new, low-defect grains form and grow. Above 298~\textdegree C, the decrease in KAM continues but at a lower rate. This behaviour is likely associated with the final stages of recrystallization together with the onset of grain growth, during which the remaining stored energy is reduced more gradually. Overall, the evolution of the normalized average KAM closely follows the trend observed in the hardness measurements (Figure \ref{fig:05}). In both cases, an initial region of moderate softening is followed by a pronounced decrease, supporting the interpretation that recovery dominates at lower temperatures, whereas recrystallization is the primary softening mechanism at higher temperatures. It should be emphasized that KAM alone cannot differentiate between recovery, recrystallization nucleation, and the early stages of recrystallisation, as all three mechanisms lead to a moderate reduction in local misorientation. Consequently, the assignment of specific transformation stages requires additional microstructural evidence from the in situ EBSD observations.

For a more detailed analysis of the mechanisms active during annealing, the total lengths of low-angle grain boundaries (LAGBs) and high-angle grain boundaries (HAGBs) were plotted as a function of temperature (Figure \ref{fig:Moritz_3}). LAGBs were defined by misorientation angles between 0.5° and 5°, whereas HAGBs were defined by misorientation angles greater than 5°. The lower threshold of 0.5° was chosen as three times the orientation noise measured in the noisiest in situ EBSD dataset. Additional details on this analysis are provided in Figure S3 of the supplemental information. The upper threshold was chosen so that the grain boundary map accurately reproduces the visible boundaries in the CI and IQ maps in the recrystallized state, even though it is below the value of 15° specified in the classical definition for high angle boundaries. Up to approximately 198~\textdegree C, neither the LAGB nor the HAGB length exhibits a significant change. Between 198~\textdegree C and 265~\textdegree C, however, the total LAGB length decreases markedly, while the HAGB length shows only a slight reduction. The decrease in LAGB length is characteristic of recovery, during which dislocations are annihilated and subgrain structures evolve through boundary migration, causing some subgrains to grow at the expense of others and thereby reducing the overall LAGB length. During this stage, a limited number of grains also begin to grow and act as potential nuclei for recrystallization, which is consistent with the relatively small change in HAGB length. Above 265~\textdegree C, both the LAGB and HAGB lengths decrease rapidly. This temperature range corresponds to the main recrystallisation stage, during which newly formed grains grow quickly at the expense of the deformed microstructure. Recrystallization continues until approximately 298~\textdegree C, where the evolution of the grain boundary lengths changes again. Beyond this temperature, both the LAGB and HAGB lengths decrease at similar rates. This behaviour most likely reflects a reduction in the driving force for recrystallisation and a transition toward grain growth or collective coarsening of the remaining subgrains and grains. Overall, the evolution of the grain boundary lengths closely follows the trend observed for the normalized average KAM, while providing additional insight into the mechanisms governing the different stages of microstructural evolution.

To further investigate the microstructural evolution during annealing, the misorientation distributions of the LAGB and HAGB segments were analysed for selected temperatures and are shown in Figure \ref{fig:Moritz_3}C and \ref{fig:Moritz_3}D, respectively. For the LAGBs, a pronounced evolution of the distribution is observed with increasing temperature. The initially broad distribution becomes increasingly concentrated at lower misorientation angles, while the overall number of LAGB segments decreases substantially. The reduction in boundary segment frequency between the green and orange histograms is consistent with recovery, during which dislocations are annihilated and subgrain boundaries are eliminated. A much more pronounced change occurs between the orange and dark blue histograms, indicating the onset and progression of recrystallization. During this stage, the remaining deformed subgrain structure is rapidly consumed by growing recrystallised grains, leading to a strong reduction in the LAGB population.

In contrast, the HAGB misorientation distributions exhibit a markedly different behaviour. The overall shape of the distribution remains largely unchanged throughout the annealing process, suggesting that no preferred HAGB misorientation range is selectively formed or removed. However, a significant reduction in the total number of HAGB segments is observed once temperatures above approximately 265~\textdegree C are reached. This decrease coincides with the recrystallisation regime identified from the KAM and grain boundary length analyses and reflects the consumption of grain boundaries as grains grow and coarsen. At higher temperatures, only a more gradual reduction in HAGB frequency is observed, which is attributed primarily to grain growth after recrystallization has largely been completed. It is worth emphasisng that a comparable analysis could not be performed for the experiment conducted at 10~\textdegree C/min due to the lower temporal resolution and the increased thermal drift during acquisition, both of which reduce the reliability of the grain boundary characterization.

\subsection{Implications of early recrystallisation for space applications}
\label{sec:dis:implications}

\noindent Radiation-damage suppression in UFG and nanocrystalline metals and alloys has been a topic of intense investigation in the past years \cite{el2014ultrafine,el2014situ,el2017role,el2018nanohardness,zhang2018radiation,el2019outstanding,barr2019interplay,el2020revealing,el2020situ,el2020temperature,el2021helium,willenshofer2022radiation,el2023quinary}. Behind this idea lies the fact that the extraordinary number of grain boundaries (GBs) act as major and strong sinks for induced radiation defects at the level of the grain microstructure: a theoretical model developed in late 1960s by Brailsford, Bullough, Hayns and co-workers \cite{brailsford1972rate,brailsford1976diffusion,brailsford1976point,bullough1980sink}. Considering an isolated spherical-shaped grain with radius $R$, these authors derived an equation for GB sink strength considering rate equations of defect generation and diffusion out of the grain. Such an expression is written as \cite{bullough1980sink,aradi2020damage}:

\begin{equation}
k_{\mathrm{gb}}^{2} = k_{\mathrm{sc}}^{2}
\left[
\frac{k_{\mathrm{sc}} R \, \coth(k_{\mathrm{sc}} R) - 1}
{1 + \dfrac{k_{\mathrm{sc}}^{2} R^{2}}{3} - k_{\mathrm{sc}} R \, \coth(k_{\mathrm{sc}} R)}
\right]
\label{eq:kgb}
\end{equation}

In equation \ref{eq:kgb}, the parameters $k_{\mathrm{gb}}$ and $k_{\mathrm{sc}}$ are the effective surface sink strength and internal sink strength, respectively. In the asymptotic limit comprising large grain sizes, \textit{i.e.}, $k_{\mathrm{sc}}R \rightarrow \infty$, the grain boundary sink strength is:

\begin{equation}
k_{\mathrm{gb}}^{2} \approx \frac{3 k_{\mathrm{sc}}}{R}
\label{eq:kgb-approx}
\end{equation}

When the metal or alloy exhibit GB sizes in the nanoscale regime, the sink strength shows a lower asymptotic limit, \textit{i.e.} for $k_{\mathrm{sc}}R \rightarrow 0$:

\begin{equation}
k_{\mathrm{gb}}^{2} \approx \frac{15}{R^{2}}
\label{eq:kgb-approx2}
\end{equation}

Here, one can clearly see the advantage of metals and alloys with GB sizes confined within the nanocrystalline and/or UFG regime: the GB sink strength is inversely proportional to the average grain radius, therefore, smaller grains will exhibit a quadratic-power sink strength compared with micro- and mesoscale-grained alloys. Valiev and Langdon \cite{valiev2006principles} noted that UFG aluminium alloys can recrystallise at temperatures around 200~\textdegree C, but a precise experimental method to determine the onset temperature of grain-boundary instabilities in such severely deformed alloys had not yet been established. This is relevant to the use of UFG aluminium alloys in space structures: if the microstructure recrystallises under the Sun's thermal irradiation, the resulting coarser grains would degrade both mechanical properties and irradiation resistance.

Here in our work, we have so far shown that \textit{in situ} EBSD heating provides significant advantages over \textit{in situ} TEM heating, considering the study of recrystallisation mechanisms in UFG aluminium alloys. Specially for the UFG AA6061 alloy system, we have identified that recrystallisation phenomena is preceded by (i) recovery and nucleation of new grains, (ii) recrystallisation \textit{per se}, and (iii) terminal grain growth; these processes take place at specific temperatures or intervals: 198~\textdegree C, 198-265~\textdegree C and 298~\textdegree C, respectively. To translate these microstructural observations into a predictable quantitative measure of radiation tolerance, we evaluated the grain-boundary sink strength of the UFG AA6061 alloy as a function of temperature via the application of Eq.~\ref{eq:kgb}, and using the average grain sizes measured \textit{in situ} EBSD during heating (Fig.~\ref{fig:Moritz_3}).

\begin{figure}[ht!]
\centering
\includegraphics[width=1\textwidth]{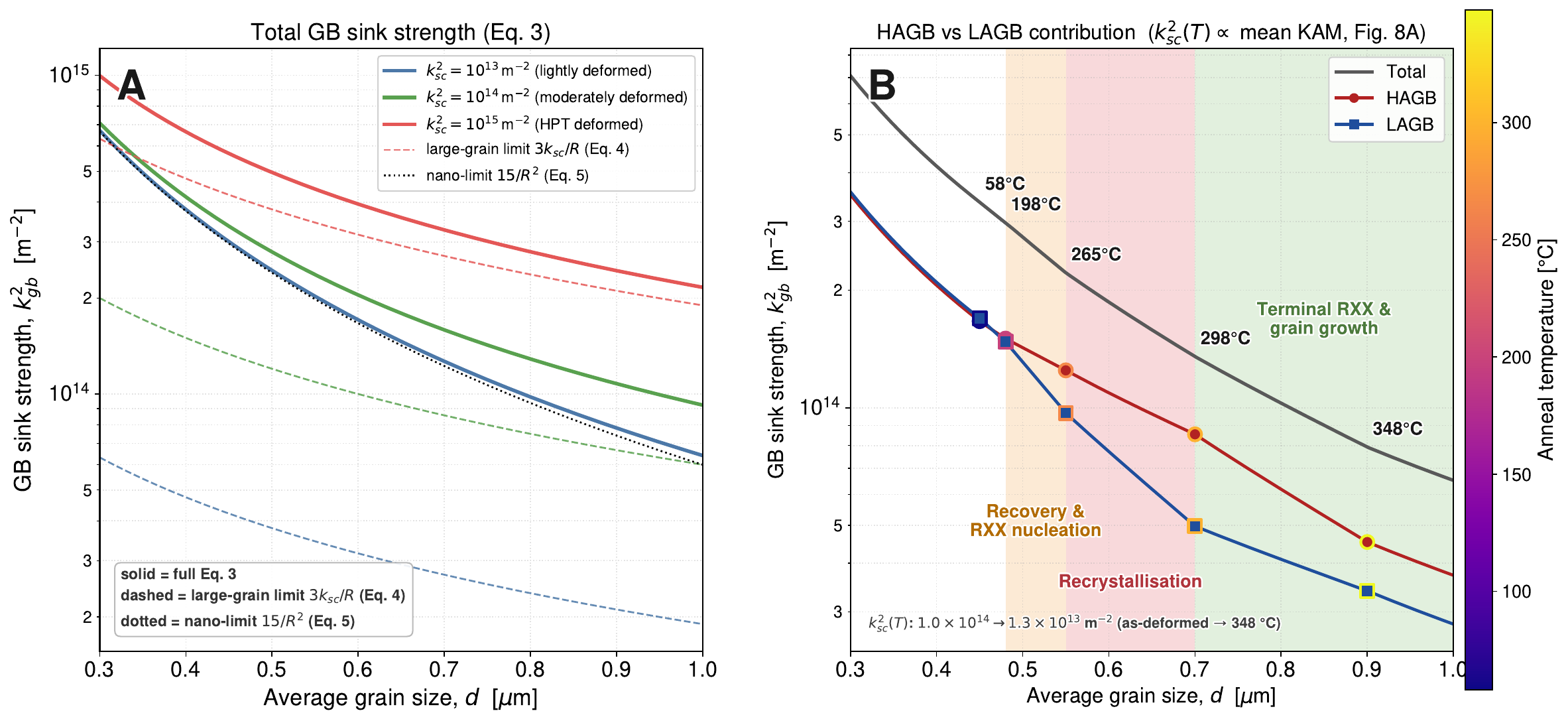}
\caption{\textbf{Grain-boundary sink strength of UFG AA6061 across the annealing sequence.} \textbf{A} Total $k_{\mathrm{gb}}^{2}$ (Eq.~\ref{eq:kgb}) versus average grain size for $k_{\mathrm{sc}}^{2}=10^{13},10^{14},10^{15}~\mathrm{m}^{-2}$ (lightly, moderately and HPT-deformed); solid: full expression, dashed: large-grain limit (Eq.~\ref{eq:kgb-approx}), dotted: nanoscale limit (Eq.~\ref{eq:kgb-approx2}). \textbf{B} Temperature-resolved $k_{\mathrm{gb}}^{2}$ with $k_{\mathrm{sc}}$ scaled by the mean KAM (Eq.~\ref{eq:mean_KAM}), partitioned into HAGB and LAGB by their relative lengths. Markers are the five EBSD temperatures (58--348~\textdegree C, colour bar); bands denote the recovery \& RXX nucleation, recrystallisation, and terminal RXX \& grain-growth regimes.}
\label{fig:Mattzi_4}
\end{figure}

 Figure~\ref{fig:Mattzi_4}A presents $k_{\mathrm{gb}}^{2}$ as a function of the average grain size for a range of internal sink strengths $k_{\mathrm{sc}}^{2}$ spanning the lightly deformed to the severely (HPT) deformed condition; for the sub-micrometre grains examined here the full expression lies well above its large-grain asymptote (Eq.~\ref{eq:kgb-approx}) and approaches the nanoscale limit (Eq.~\ref{eq:kgb-approx2}), confirming that the ultrafine-grained structure operates in the regime where the sink strength scales quadratically with $R^{-1}$. Because the internal sink strength of UFG metals \& alloys is governed primarily by the stored dislocation density -- which is progressively removed during recovery and recrystallisation -- $k_{\mathrm{sc}}$ cannot be treated as constant throughout the anneal, as it is in the original isothermal formulation of Brailsford, Bullough and Hayns \cite{brailsford1972rate,brailsford1976diffusion,brailsford1976point,bullough1980sink}. We, therefore, extend their model by allowing the internal sink strength to evolve with temperature, using the mean KAM as a proxy for the stored dislocation content, \textit{i.e.}:

\begin{equation}
k_{\mathrm{sc}}^{2}(T) = k_{\mathrm{sc},0}^{2}\, \frac{\overline{\mathrm{KAM}}(T)}{\overline{\mathrm{KAM}}_{0}}
\label{eq:ksc-kam}
\end{equation}

Where $\overline{\mathrm{KAM}}(T)$ is the measured mean KAM at temperature $T$ (Fig.~\ref{fig:Moritz_3}A) and $k_{\mathrm{sc},0}^{2}$ and $\overline{\mathrm{KAM}}_{0}$ are the values in the as-deformed HPT reference state. Combining Eqs.~\ref{eq:kgb} and~\ref{eq:ksc-kam} yields a temperature-resolved internal sink strength that couples the two degradation channels active during annealing: grain coarsening ($R$ increasing) and dislocation annihilation ($k_{\mathrm{sc}}$ decreasing).

Figure~\ref{fig:Mattzi_4}B shows the resulting evolution of $k_{\mathrm{gb}}^{2}$, with the total sink strength additionally partitioned between high- and low-angle boundaries according to their measured relative lengths (Fig.~\ref{fig:Moritz_3}B). As the microstructure passes from the HPT as-deformed state (58~\textdegree C) to terminal grain growth (348~\textdegree C), the internal sink strength falls from $k_{\mathrm{sc}}^{2}\approx1.0\times10^{14}$ to $1.3\times10^{13}~\mathrm{m}^{-2}$ and the total grain-boundary sink strength drops by roughly a factor of four, from $\sim\!3.4\times10^{14}$ to $\sim\!7.9\times10^{13}~\mathrm{m}^{-2}$; because grain coarsening and dislocation annihilation act in the same direction, the decline is markedly steeper than that predicted from grain coarsening alone (\textit{i.e.}, the original theory from Bullough \textit{et al.} \cite{brailsford1972rate,brailsford1976diffusion,brailsford1976point,bullough1980sink}). The steepest loss coincides with the recovery, nucleation and recrystallisation intervals (198--298~\textdegree C), and the boundary character controlling the sink strength shifts over the same window: low- and high-angle boundaries contribute almost equally in the deformed state, but as the sub-boundary network is consumed the high-angle boundaries come to dominate the residual sink strength. The practical implication for space applications is direct: since the onset of these instabilities occurs at 198~\textdegree C -- the upper bound of the LEO service temperature -- the grain-boundary sink strength that underpins the radiation tolerance of UFG AA6061 begins to degrade precisely within the operational window, reinforcing the conclusion that the alloy in its current condition is poorly suited to prolonged solar-thermal exposure.

Considering the mechanisms of materials degradation in space, and given that aluminium alloys can reach temperatures of up to 200~\textdegree C when exposed to solar irradiation under LEO conditions \cite{tunes2025future}, the onset of grain microstructure instabilities in the UFG AA6061 alloy investigated in this work occurs at 198~\textdegree C -- almost exactly the upper bound of the expected service temperature. Because recrystallisation and the associated grain coarsening would set in within this operational window, degrading both the mechanical properties and the irradiation resistance imparted by the ultrafine-grained structure, the suitability of this alloy as a candidate material for space applications is severely limited.

\section{Conclusions}
\label{sec:conclusions}
\noindent The thermal stability and recrystallisation behaviour of an ultrafine-grained AA6061 (Al--Mg--Si) alloy, produced by high-pressure torsion, were investigated using two complementary \textit{in situ} electron microscopy techniques, supported by DSC, STEM--EDX and microhardness measurements. The main conclusions are:

\begin{enumerate}[label=(\roman*),leftmargin=*]
    \item \textit{In situ} EBSD heating outperforms \textit{in situ} TEM heating for determining the onset of recrystallisation in UFG alloys. It samples $\mathcal{O}(10^3)$ grains from bulk material, is free of the thin-film effect that shifts thermal events to lower temperatures in electron-transparent lamellae, and exploits orientation contrast to resolve grain-boundary activity unambiguously.

    \item At a heating rate of 1~\textdegree C$\cdot\mathrm{min}^{-1}$, \textit{in situ} EBSD resolved the microstructural evolution into three sequential regimes: recovery and nucleation from $\sim$198~\textdegree C, recrystallisation between $\sim$198 and 265~\textdegree C, and grain growth above $\sim$298~\textdegree C. Combined KAM and grain-boundary-length analyses corroborated this sequence, with recovery-driven LAGB annihilation preceding HAGB consumption during recrystallisation.

    \item DSC and STEM--EDX revealed that the UFG condition suppresses GP-zones formation and shifts precipitation to lower temperatures relative to the coarse-grained alloy ($\beta^{\prime}$ at 225~\textdegree C, $\beta$-Mg$_2$Si at 318~\textdegree C), consistent with accelerated diffusion along the high-density grain-boundary network. Crucially, precipitation neither retards recrystallisation nor restores strength once the ultrafine-grained structure is lost, as confirmed by the monotonic decay in microhardness.

    \item The microhardness response mirrors the \textit{in situ} microstructural evolution: the steep drop in hardness between the as-HPT state and $\sim$242~\textdegree C coincides with the recovery-to-recrystallisation transition tracked by the normalised KAM, confirming that the strength imparted by high-pressure torsion is forfeited as the ultrafine-grained structure recrystallises, while the negligible hardening at 318~\textdegree C shows that $\beta$-phase precipitation offers no (further) compensating strengthening.

    \item Applying the Brailsford--Bullough--Hayns sink-strength model to the
    \textit{in situ} grain sizes places the as-HPT structure in the $R^{-2}$ regime
    (Eq.~\ref{eq:kgb-approx2}), well above the large-grain limit. Extending the
    isothermal theory by scaling the internal sink strength with the measured mean
    KAM, $k_{\mathrm{sc}}^{2}(T)\propto\overline{\mathrm{KAM}}(T)$
    (Eq.~\ref{eq:ksc-kam}), shows that grain coarsening and dislocation annihilation
    together lower the total grain-boundary sink strength by about a factor of four
    between the as-deformed state and 348~\textdegree C, most steeply across the
    198--298~\textdegree C recrystallisation window, with high-angle boundaries
    progressively dominating as the sub-boundary network is consumed.

    \item The early onset of grain-microstructure instabilities ($\sim$198~\textdegree C) coincides with the $\sim$200~\textdegree C that aluminium alloys can reach under solar irradiation in low-Earth orbit. Consequently, this UFG AA6061 alloy would recrystallise within its intended service window, forfeiting the mechanical strength and grain-boundary-mediated radiation tolerance that motivate its use, which severely limits its suitability as a candidate space material.
\end{enumerate}

Considering general metallurgy and materials science applicability, \textit{in situ} EBSD heating emerges as a powerful \textit{in operando} tool for the accurate, bulk-representative determination of microstructural instabilities in candidate space materials. Moreover, by revisiting the Brailsford--Bullough--Hayns sink-strength theory and extending it with a temperature-dependent internal sink strength inferred from the measured KAM, we have shown that the grain-boundary sink strength of the UFG structure is not fixed but decays as recovery and recrystallisation proceed. Building on this, the development of UFG aluminium alloys with recrystallisation temperatures exceeding the thermal limits of the space environment is identified as a priority for future research; equally, future work should better resolve the distinct roles of low- and high-angle grain boundaries in the overall sink strength under coupled irradiation and thermal conditions, \textit{i.e.}, in environments that provide a driving force for the initial ultrafine-grained microstructure to evolve.

\section*{Supplemental material} 
\label{extra:supplemental}
\noindent In the Mendeley dataset linked to our paper (\href{https://doi.org/10.17632/6nf9tyvgmh.4}{10.17632/6nf9tyvgmh.4}), the reader will find both \textit{in situ} TEM heating and \textit{in situ} EBSD heating videos from the experiments reported in this work:

\begin{itemize}
    \item In situ EBSD heating IPF CI - UFG AA6061 alloy - Exp 1.avi
    \item In situ EBSD heating IPF CI - UFG AA6061 alloy - Exp 2.avi
    \item In situ TEM heating - UFG AA6061 alloy - Reduced 184pc.mp4
\end{itemize}

\section*{Acknowledgement(s)}
\label{extra:acknown}
\noindent MAT and SG would like to thank Mr. Matthias Honner, Ms. Nadine Tatzreiter, and Mr. Bogdan Malysh for support with sample preparation. MT and SM would like to thank Ms. Arnela Blažević and Ms. Anita Rossmann-Perner for EBSD sample preparation. MAT, TMK and SP acknowledges funding from Austrian Research Promotion Agency (FFG) in the project 3DnanoAnalytics (FFG-No 858040). MT and SM acknowledge the Graz Centre for Electron Microscopy (ZFE) for providing the infrastructure for the (\textit{in situ}) EBSD experiments. MAT and MT would like to thank Dr. Irmgard Weißensteiner (Montanuniversität Leoben) for reading the previous version of this manuscript. MAT would like to thank Mr. Martin Hasenburger for reading the last version of the manuscript and provide valuable insights and review.

\section*{Conflict of interest}
\label{extra:conflict}
\noindent The authors declare no conflict interest.

\section*{CRediT author statement}
\label{extra:CREDIT}
\noindent MT performed experiments, wrote the initial draft with MAT, analysed and processed the data. SG and PW synthesized the alloy. TMK contributed to the \textit{in situ} TEM experiments. SM mentored MT and performed EBSD evaluation. SP provided funding for the execution of the TEM experiments and the alloy preparation. MAT conceptualisation of the study, mentored SG, wrote the initial draft with MT, analysed and processed the data. FF provided mentorship, experimental expertise and guidance with microhardness. All authors participated in reviewing the manuscript draft.

\section*{Data repository and availability}
\label{extra:data}
\noindent We have uploaded raw and processed data and the necessary steps to reproduce the results in this paper in a Mendeley Dataset. This can be access at: M. Theissing \textit{et al.} (2025): ``Data for: Thermal stability and recrystallisation dynamics of ultrafine-grained aluminium alloys studied with in situ EBSD and TEM'', Mendeley Data, version 4, doi: \href{https://doi.org/10.17632/6nf9tyvgmh.4}{10.17632/6nf9tyvgmh.4}

\bibliographystyle{elsarticle-num.bst}
\bibliography{bibdata_updated.bib}
\end{document}